# Visualization study of counterflow in superfluid helium-4 using metastable helium molecules

Wei Guo[1], Sidney B. Cahn[1], James A. Nikkel[1], William F. Vinen[2] & Daniel N. McKinsey[1]

[1]Physics Department, Yale University, New Haven, CT 06520, USA

[2]School of Physics and Astronomy, University of Birmingham, Birmingham B15 2TT, United Kingdom

**Heat is carried in superfluid $^4$He by the motion of the normal fluid[1], a counterflowing superfluid component serving to eliminate any net mass flow. It has been known for many years that above a critical heat current the superfluid component in this counterflow becomes turbulent. This turbulence takes the form of a disorganized tangle of quantized vortex lines and is maintained by the relative motion of the two fluids[2-3]. It has been suspected that the normal fluid may also become turbulent[4], but experimental verification is difficult without a technique for visualizing the flow. Here we report a series of visualization studies on the normal-fluid component in a thermal counterflow performed by imaging the motion of seeded metastable helium molecules using a laser-induced-fluorescence technique[5-8]. We present evidence that the flow of the normal fluid is indeed turbulent, at least at relatively large velocities. Thermal counterflow in which both components are turbulent presents us with a new and theoretically challenging type of turbulent behaviour.**

The superfluid phase of liquid $^4$He exhibits two-fluid behaviour[1]: a normal fluid, carrying all the thermal energy, coexists with a superfluid component. The proportion of superfluid falls from unity to zero as the temperature, $T$, rises from zero to the λ-transition[1]. In a thermal counterflow, the normal-fluid velocity, $v_n$, is related to the heat flux, $q$, by

$$q = \rho s T v_n , \tag{1}$$

where $\rho$ is the total density and $s$ is the entropy per unit mass[1]. Above a certain critical heat flux, superflow is known to become dissipative (quantum turbulence)[2-3]. A mutual friction force between the two fluids arises through the interaction between quantized vortices and the normal fluid[9]. This type of quantum turbulence is maintained by the relative motion of the two fluids, through processes that are reasonably well understood[2-3]. Features in the observed relation between vortex density and heat flux suggest that the normal fluid may also become turbulent, and mutual friction has been shown theoretically to induce an instability in the laminar flow of the normal fluid[4]. Satisfactory evidence for such normal-fluid turbulence can come only from a visualization of the normal-fluid flow. Techniques for such visualization have recently started to be developed. Turbulence in both fluids has been observed in other types of flow[10] in liquid helium, such as that behind a moving grid[11]. But in those cases the two fluids are not forced to have any relative motion, and behave like a single classical fluid, exhibiting a Kolmogorov energy spectrum[1]. Simultaneous turbulence in both fluids in a counterflow must be different, and it would be a type of turbulence that is new to physics.

Past experiments on the visualization of thermal counterflow have used micron-sized tracer particles formed from polymer spheres or solid hydrogen, and they have been based on either Particle Image Velocimetry[12-13] (PIV) or particle tracking techniques[14-16]. The PIV data obtained at large heat fluxes are hard to interpret since micron-sized particles can be trapped on the quantized vortex lines. However, particle tracking has yielded very interesting results, demonstrating that the normal fluid does move according to equation (1), and that information about the dynamical behaviour of the quantized vortices can be obtained. However, these particle tracking experiments have been confined to small heat currents, at which there is no clear indication of any normal-fluid turbulence. The results at larger heat currents, when vortex density is high, may be harder to interpret[17].

Recently, we have demonstrated that metastable $He_2^*$ triplet molecules, with a radiative lifetime of about 13 s in liquid helium[18], can be imaged using a laser-induced-fluorescence technique[5-8] (see Supplementary Information Sec. 1 and Fig. S1). These molecules can act as tracers that follow the motion of the normal fluid. Their small size (~ 1 nm) means that they are not trapped by vortices at temperatures above 1 K[19], and scattering by vortices is likely to have a negligible effect. In this paper, we report the results of visualization studies on the normal fluid in a thermal counterflow using these molecules.

In one experiment, we used a molecule tagging technique[8] to visualize the velocity profile of the normal fluid. The experimental set-up is shown in Fig. 1a. A counterflow channel, consisting of a square borosilicate glass tube (8 cm long, 5 mm inner side width) with one end sealed to a stainless steel flange (see Fig. 1b), was installed in a helium cell. Two sharp tungsten needles were clamped inside the steel flange and surrounded by a copper ring electrode as a field emission source. When a negative voltage with amplitude greater than the emission threshold (roughly -500 V) was applied to the needles, electrons were emitted from each needle and moved into the copper electrode. Metastable helium molecules were produced near the needle apexes[7,20] with a rate on the order of $10^9$ $s^{-1}$. A heater installed inside the steel flange served to generate a counterflow in the channel. A focused pump laser pulse (see Supplementary Information Sec. 2) at 910 nm was used to drive the $He_2^*$ molecules along the beam path into the first vibrational level of the triplet ground state (see Fig. 2a). At a given delay time, an expanded probe laser pulse at 925 nm was then used to selectively image the vibrationally excited line of molecules by driving them to an excited electronic state and inducing 640 nm fluorescent light. In the absence of a heat current, we saw no molecules in the laser region while the field emission source was on. As we turned on the heater, helium molecules drifted with the normal fluid to the laser region and were imaged. At low heat fluxes, however, many molecules decayed radiatively before they could reach the observation region. The flow of the normal fluid was studied at heat fluxes that ranged from 160 to 1,000 mW/cm$^2$, above the onset heat flux for quantum turbulence (about 20 mW/cm$^2$) observed in channels with similar geometry[21-22].

The fluorescent light from the line of excited molecules was recorded by an intensified CCD camera (see Supplementary Information Sec. 3). To achieve high image quality at each given pump-probe delay time, up to 40 images were superimposed and an averaged line profile was obtained. Typical summed images showing the motion of a line of tagged molecules at a heat flux of 640 mW/cm$^2$ are shown in Fig. 2b in false color. For all studied heat fluxes, an initially straight line of molecules remains straight as it drifts, indicating a flat normal-fluid velocity profile across the channel. For each image, we integrate the fluorescent signal in each pixel along the x-axis for a given y (see Fig. 2a for the axes) and plot the summed signal as a function of y to show the integrated cross-section profile of the tagged line. An example is shown in Fig. 2c. The resulting line profile can be fit well using a Lorentzian function, matching the Lorentzian profile of the pump laser beam. The center position and the mean width of a molecule line can be determined. The normal-fluid velocity at a given heat flux is obtained by a linear fit of the center position as a function of drift time and is shown in Fig. 3a. The solid line in Fig. 3a shows the calculated normal-fluid velocity based on equation (1). The good agreement between our data and the theory demonstrates that the fluorescing helium molecules do indeed track the normal fluid. An approximately flat normal-fluid velocity profile in counterflow was also found by Awschalom *et al.*[23] using a cluster of electrons as a tracer. The observation in our experiment agrees with their findings, with a simplification in interpretation due to the fact that unlike electrons, helium molecules do not repel each other and interact less strongly with quantized vortices.

There remains the question of whether the flow of the normal fluid is turbulent. Two effects can cause a flattening of the normal-fluid velocity profile: turbulence, and non-linear mutual friction acting on a laminar normal-fluid flow. It is therefore not clear whether the flat profile is clear evidence for turbulence. Better evidence comes from the observed broadening of the line of molecules with increasing time. We analysed this broadening at a heat flux of 277 mW/cm$^2$, although similar broadening is seen at all heat fluxes that we have studied. In Fig. 3b we plot the mean square width $\langle w^2(t) \rangle$ of the line as a function of time. This increase occurs much more rapidly than can be accounted for by ordinary diffusion of the molecules at the temperature (1.95K) of this experiment[5]. We deduce that it is caused by turbulent diffusion[24-25], thus confirming that the flow of the normal fluid is turbulent.

The quantity $\left\langle w^2(t)\right\rangle$ is a measure of the time dependence of the separation of two particles in the turbulent flow[24-25]. Its rate of increase at any instant is governed largely by eddies of size equal to this separation. The time dependence of $\left\langle w^2(t)\right\rangle$ depends on the energy spectrum, $E(k)$, characterising the turbulence. Predictions exist for the case of a Kolmogorov spectrum, $E(k) \sim k^{-5/3}$. For times that are large compared with the turnover time of the relevant eddies, and for separations that are within an inertial range, the Kolmogorov spectrum leads to[24] $\left\langle w^2(t)\right\rangle \sim t^3$. Our data suggest that the exponent of $t$ for normal-fluid turbulence in a counterflow is close to 4, which means the turbulent energy spectrum is not of the Kolmogorov form. Further experiments will be conducted, and we are developing a theory that will relate the exponent of $t$ to the exponent of $k$ in the energy spectrum. Knowledge of the turbulent energy spectrum in the normal fluid in a counterflow will be valuable to understand this new form of turbulence. We emphasize that the heat fluxes at which we see evidence for turbulence in the normal fluid are significantly larger than those studied by Paoletti *et al*[14].

To measure the normal-fluid velocity at small heat fluxes, we used a cluster tracking technique[8] in a narrower counterflow channel: a square glass tube 6 cm long with 2 mm inner side width. Two tungsten needles (0.1 mm in diameter) were precisely aligned along the central axis of the channel with a gap between them of about 1.5 mm (see Fig. 1c). When a negative voltage pulse was applied to the cathode needle, a small cluster of helium molecules was created near its apex with an initial diameter in the range of 0.5 mm to 0.8 mm. This molecule cluster serves as a single tracer[8]. A pulsed laser at 905 nm (repetition rate 500 Hz) illuminated the needle apexes (see Fig. 4a) to drive the molecules to produce fluorescent light through a cycling transition[5-8]. Typical images of a molecule cluster at a heat flux of 119 mW/cm$^2$ are shown in Fig. 4b in false color. The normal-fluid velocity is obtained by a linear fit of the cluster center as a function of drift time and is plotted in Fig. 4c. Our data again agree well with theory, except below a transition heat flux at about 50 mW/cm$^2$, which may correspond to the onset of quantum turbulence[21-22, 26]. It is likely that, in the small heat flux regime, both the superflow and the normal-fluid flow are laminar. A molecule cluster moving along the central axis should have a drift velocity roughly twice as large as the mean velocity given by equation (1) due to a Poiseuille (parabolic) profile[1] of the normal fluid[27]. Indeed, if we divide the measured velocities in the small heat flux regime by a factor of 2, the data fall back to the theory line (see the blue crosses in the zoom-in inset in Fig. 4c). Above the transition heat flux, when the superfluid (perhaps also the normal fluid) becomes turbulent, the normal-fluid profile is flattened[4] and a molecule cluster then moves at a velocity equal to the mean velocity.

More sophisticated studies on the normal-fluid turbulence can be conducted using a well-focused femtosecond laser to create a thin line of molecules (through laser-field ionization[28]). The line of molecules can be created inside the observation region, and hence can be imaged at any heat flux. By imaging molecules in the vibrational ground state through a cycling transition, high quality single-shot images of individual lines can be taken. The line distortion will provide us information on the structure factors[1] of the normal-fluid turbulence.

**Author contributions**

W.G. designed and performed the reported studies. D.N.M. supervised the research project and guided the manuscript editing throughout. W.F.V. provided theory assistance in data analysis and interpretation. S.B.C. and J.A.N. assisted in the experiment design. W.G. wrote the manuscript. All authors discussed the results and commented on the manuscript.

**Competing financial interests**

The authors declare that they have no competing financial interests.

1. Landau, L.D. & Lifshitz, E.M. *Fluid Mechanics* (Pergamon Press, Oxford, UK, 1987).
2. Vinen, W.F. Mutual friction in a heat current in liquid helium II. III. Theory of the mutual friction. *Proc. R. Soc. London. Ser. A* **242,** 493-515 (1957).
3. Schwarz, K.W. Three-dimensional vortex dynamics in superfluid $^4$He: Homogeneous superfluid turbulence. *Phys. Rev. B* **38,** 2398-2417 (1988).
4. Melotte, D.J. & Barenghi, C.F. Transition to normal fluid turbulence in helium II. *Phy. Rev. Lett.* **80,** 4181-4184 (1998).
5. McKinsey, D.N., Lippincott, W.H., Nikkel J.A. & Rellegert, W.G. Trace detection of metastable helium molecules in superfluid helium by laser-induced fluorescence. *Phys. Rev. Lett.* **95,** 111101/1-4 (2005).

6. Rellergert, W.G. *et al*. Detection and imaging of $He_2$ molecules in superfluid helium. *Phys. Rev. Lett*. **100,** 025301/1-4 (2008).

7. Guo, W., Wright, J.D., Cahn, S.B., Nikkel, J.A. & McKinsey, D.N. Studying the normal-fluid flow in helium-II using metastable helium molecules. *J. Low Temp. Phys.* **158,** 346-352 (2010).

8. Guo, W., Wright, J.D., Cahn, S.B., Nikkel, J.A. & McKinsey, D.N. Metastable helium molecules as tracers in superfluid $^4$He. *Phy. Rev. Lett.* **102,** 235301/1-4 (2009).

9. Hall, H.E. & Vinen, W.F. The rotation of liquid helium II. II. The theory of mutual friction in uniformly rotating helium II. *Proc. R. Soc. London. Ser. A* **238,** 215-234 (1956).

10. Vinen, W.F. & Niemela, J.J. Quantum turbulence. *J. Low. Temp. Phys.* **128,** 167-231 (2002).

11. Stalp, S.R., Skrbek, L. & Donnelly, R.J. Decay of grid turbulence in a finite channel. *Phys. Rev. Lett.* **82,** 4831-4834 (1999).

12. Van Sciver, S.W., Fuzier, S. & Xu, T. Particle image velocimetry studies of counterflow heat transport in superfluid helium II. *J. Low Temp. Phys.* **148,** 225-233 (2007).

13. Zhang, T. & Van Sciver, S.W. Large-scale turbulent flow around a cylinder in counterflow superfluid $^4$He (He(II)). *Nat. Phys.* **1,** 36-38 (2005).

14. Paoletti, M.S., Fiorito, R.B., Sreenivasan, K.R. & Lathrop, D.P. Visualization of superfluid helium flow. *J. Phys. Soc. Jap.* **77,** 111007/1-7 (2008).

15. Bewley, G.P., Lathrop, D.P. & Sreenivasan, K.R. Superfluid helium: Visualization of quantized vortices. *Nature* **441,** 588 (2006).

16. Paoletti, M.S., Fisher, M.E., Sreenivasan, K.R. & Lathrop, D.P. Velocity statistics distinguish quantum turbulence from classical turbulence. *Phys. Rev. Lett.* **101,** 154501/1-4 (2008).

17. Kivotides, D. Motion of a spherical solid particle in thermal counterflow turbulence. *Phys. Rev. B* **77,** 174508/1-5 (2008).

18. McKinsey, D. N. *et al*. Radiative decay of the metastable $He_2\left(a^3\Sigma_u^+\right)$ molecule in liquid helium. *Phys. Rev. A* **59,** 200-204 (1999).

19. Vinen, W. F. *Progress in Low Temperature Physics* AIP Conf. Proc. No. **850,** 169 (AIP, New York, 2006).

20. Mehrotra, R., Mann, E.K. & Dahm, A.J. A study of the neutral excitation current in liquid $^4$He above 1K. *J. Low Temp. Phys.* **36,** 47-65 (1979).

21. Vinen, W.F. Mutual friction in a heat current in liquid helium II. IV. Critical heat current in wide channels. *Proc. R. Soc. London. Ser. A* **243,** 400-413 (1958).

22. Martin, K.P. & Tough, J.T. Evolution of superfluid turbulence in thermal counterflow. *Phys. Rev. B* **27,** 2788-2799 (1983).

23. Awschalom, D.D., Milliken, F.P. & Schwarz, K.W. Properties of superfluid turbulence in a large channel. *Phys. Rev. Lett.* **53,** 1372-1375 (1984).

24. Batchelor, GK. Diffusion in a field of homogeneous turbulence. II. The relative motion of particles. *Proc. Cambridge Philos. Soc*. **48,** 345–362 (1952a).

25. Sawford, B. Turbulent relative dispersion. *Annu. Rev. Fluid Mech.* **33,** 289-317 (2001).

26. Chase, C.E. Thermal conduction in liquid helium II. II. Effects of channel geometry. *Phys. Rev.* **131,** 1898-1903 (1963).

27. Spiga, M. & Morini, G.L. A symmetric solution for velocity profile in laminar flow through rectangular ducts. *Int. Comm. Heat Mass Transfer.* **21,** 469-475 (1994).

28. Benderskii, A.V., Zadoyan, R., Schwentner, N. & Apkarian, V.A. Photodynamics in superfluid helium: Femtosecond laser-induced ionization, charge recombination, and preparation of molecular Rydberg states. *J. Chem. Phys.* **110,** 1542-1557 (1999).

Figure 1: **Experimental set-up**. **a,** The equipment used in the experiment is shown schematically. Lasers pass through the helium cell, and an intensified CCD camera views the counterflow channel from the bottom. The helium cell is thermally linked to a helium bath whose temperature can be controlled. **b,** The counterflow channel used in the molecule tagging experiment is shown schematically. The square glass tube has an inner side width of 5 mm and a length of 8 cm. The pump and probe lasers illuminate the channel at about 6 cm from the tungsten needles. **c,** The counterflow channel used in the cluster tracking experiment is shown schematically. The square glass tube has an inner side width of 2 mm and a length of 6 cm. A pulsed laser at 905 nm (repetition rate 500 Hz) illuminates the channel at about 2 cm from the open end of the channel.

Figure 2: **Typical fluorescent images and graph of a line of tagged helium molecules**. **a,** An image of the square glass tube is shown together with a schematic showing the widths and the positions of the pump and the probe lasers. **b,** Typical fluorescent images of a molecule line at pump-probe delay of 0 ms, 40 ms and 80 ms are shown. The heat flux is 640 mW/cm$^2$. The temperature is 1.95 K. Each image is a superposition of 40 pump-probe trials. **c,** The black curve shows a typical integrated cross-section profile of a line of tagged molecule in arbitrary units, as discussed in the text. The heat flux is 277 mW/cm$^2$, and the pump-probe delay is 20 ms. The red line is a Lorentzian fit to the data.

Figure 3: **Experiment data obtained in the molecule tagging experiment**. **a,** The black crosses show the measured normal-fluid velocity as a function of heat flux at 1.95 K. Error bars represent the error in the linear fit to determine the normal-fluid velocity as discussed in the text. The solid line shows the calculated velocity for a given heat flux based on equation (1) in the text. **b,** The black crosses show the mean square width $\left\langle w^2(t)\right\rangle$ of a molecule line as a function of drift time in the tagging experiment at a heat flux of 277 mW/cm$^2$. The solid curve is a power law fit to the data. The best fit value of the power index is 4.03.

Figure 4: **Images and data obtained in the cluster tracking experiment**. **a,** An image of the square glass tube is shown. A pulsed laser at 905 nm illuminates the region of the needle apexes. **b,** Fluorescent images of a cluster of helium molecules taken at 0 s, 0.1 s and 0.2 s after the cluster was created are shown. The heat flux is 119 mW/cm$^2$. The temperature is 1.80 K. The exposure time for each image is 20 ms. **c,** The black crosses show the measured normal-fluid velocity as a function of heat flux at 1.80 K. Error bars represent the error in the linear fit to determine the normal-fluid velocity as discussed in the text. The red line is the theory curve based on equation (1) in the text. The blue crosses in the zoom-in inset for small heat flux regime show the corresponding measured normal-fluid velocity divided by a factor of 2.

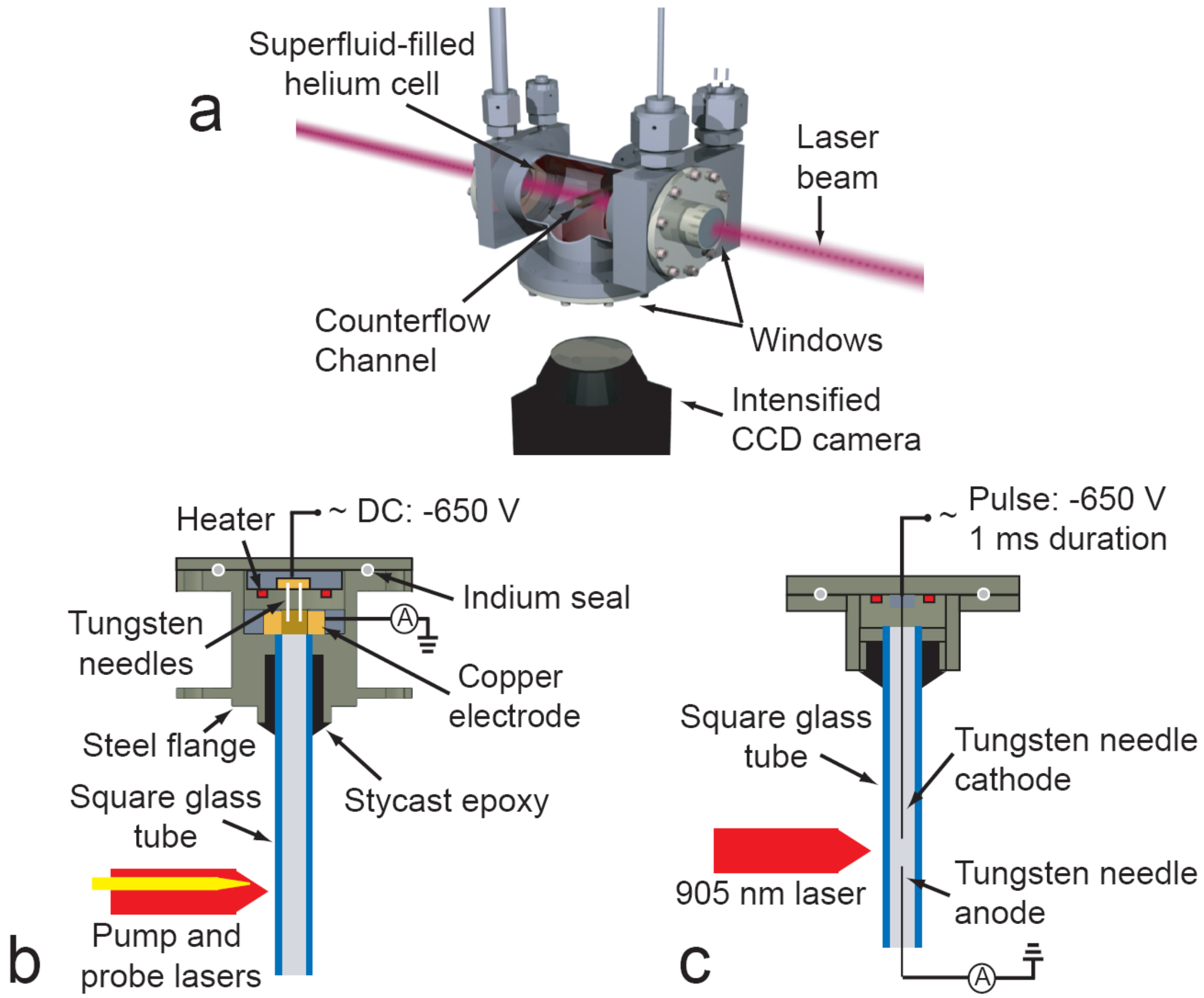

Figure 1: **Experimental set-up**. **a,** The equipment used in the experiment is shown schematically. Lasers pass through the helium cell, and an intensified CCD camera views the counterflow channel from the bottom. The helium cell is thermally linked to a helium bath whose temperature can be controlled. **b,** The counterflow channel used in the molecule tagging experiment is shown schematically. The square glass tube has an inner side width of 5 mm and a length of 8 cm. The pump and probe lasers illuminate the channel at about 6 cm from the tungsten needles. **c,** The counterflow channel used in the cluster tracking experiment is shown schematically. The square glass tube has an inner side width of 2 mm and a length of 6 cm. A pulsed laser at 905 nm (repetition rate 500 Hz) illuminates the channel at about 2 cm from the open end of the channel.

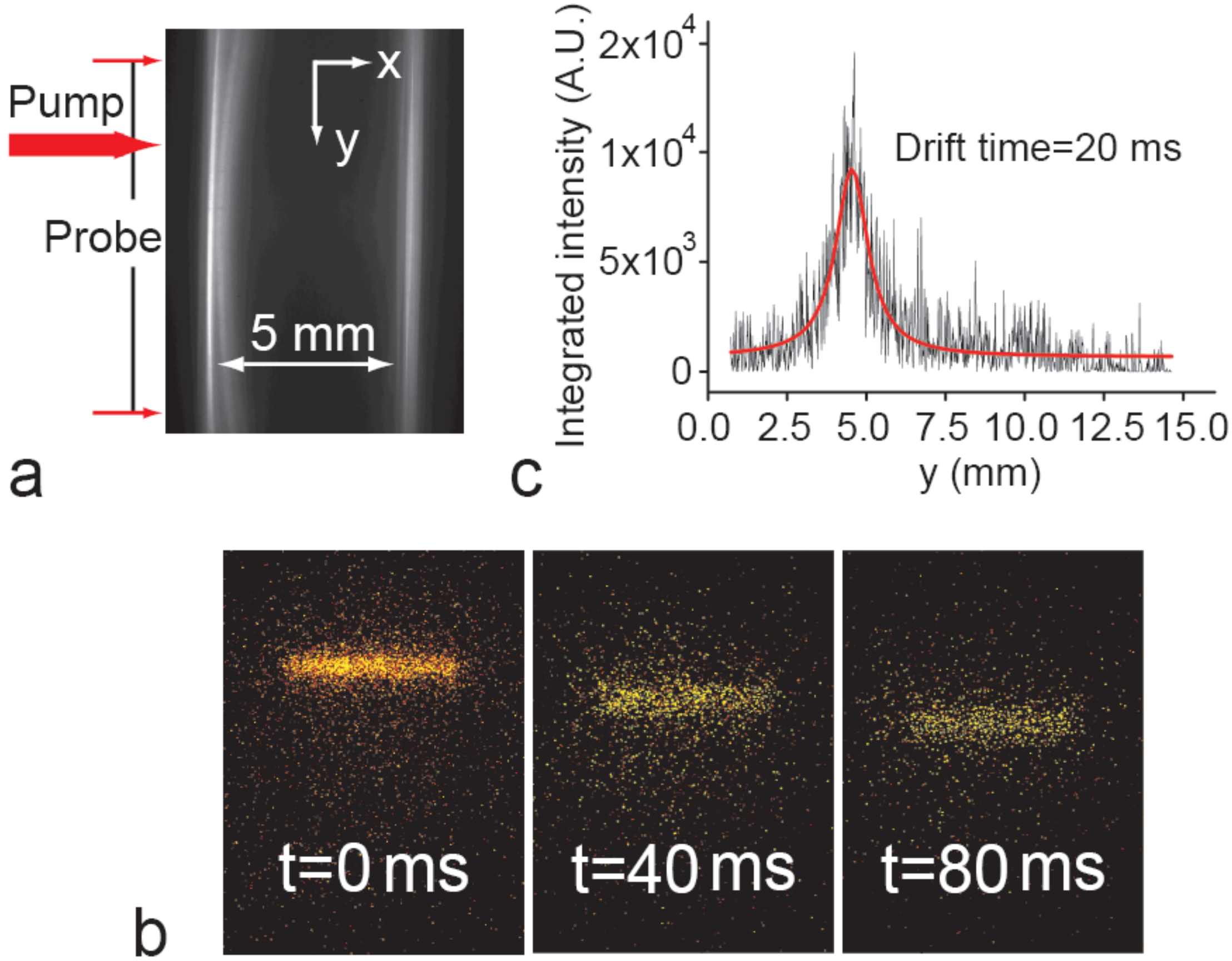

Figure 2: **Typical fluorescent images and graph of a line of tagged helium molecules**. **a,** An image of the square glass tube is shown together with a schematic showing the widths and the positions of the pump and the probe lasers. **b,** Typical fluorescent images of a molecule line at pump-probe delay of 0 ms, 40 ms and 80 ms are shown. The heat flux is 640 mW/cm$^2$. The temperature is 1.95 K. Each image is a superposition of 40 pump-probe trials. **c,** The black curve shows a typical integrated cross-section profile of a line of tagged molecule in arbitrary units, as discussed in the text. The heat flux is 277 mW/cm$^2$, and the pump-probe delay is 20 ms. The red line is a Lorentzian fit to the data.

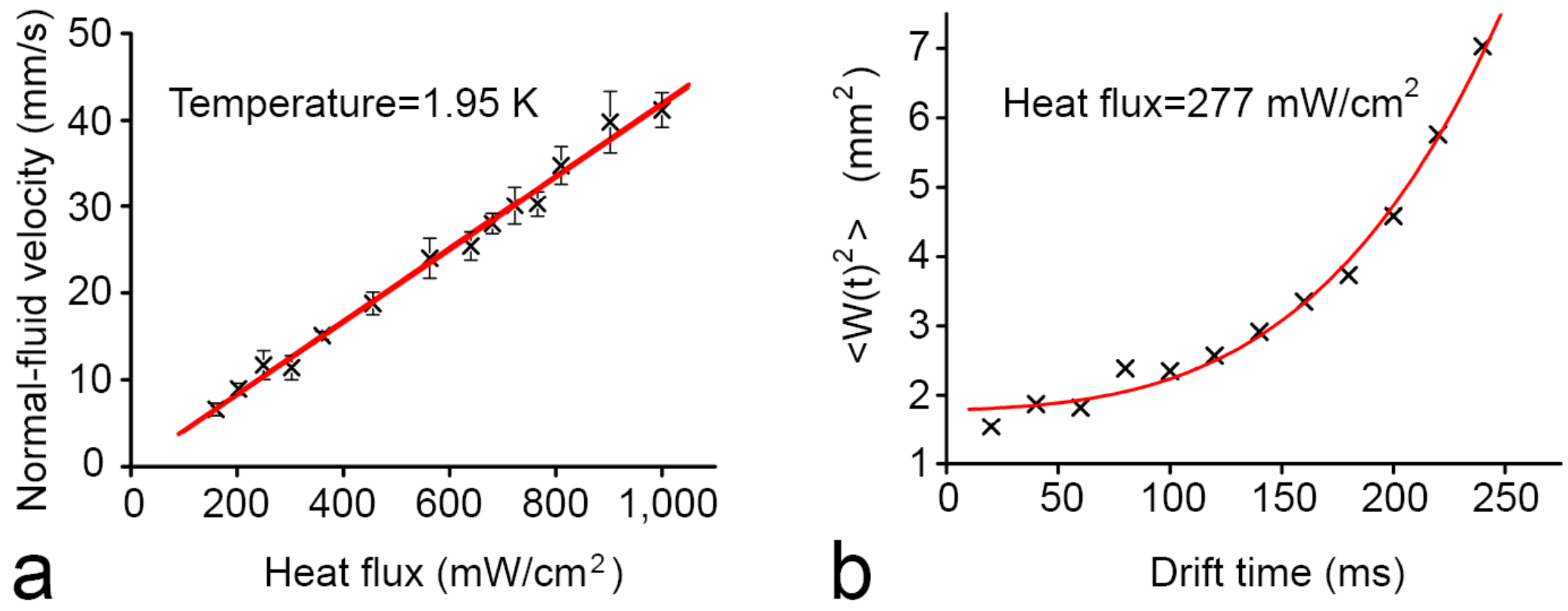

Figure 3: **Experiment data obtained in the molecule tagging experiment**. **a,** The black crosses show the measured normal-fluid velocity as a function of heat flux at 1.95 K. Error bars represent the error in the linear fit to determine the normal-fluid velocity as discussed in the text. The solid line shows the calculated velocity for a given heat flux based on equation (1) in the text. **b,** The black crosses show the mean square width $\left\langle w^2(t) \right\rangle$ of a molecule line as a function of drift time in the tagging experiment at a heat flux of 277 mW/cm$^2$. The solid curve is a power law fit to the data. The best fit value of the power index is 4.03.

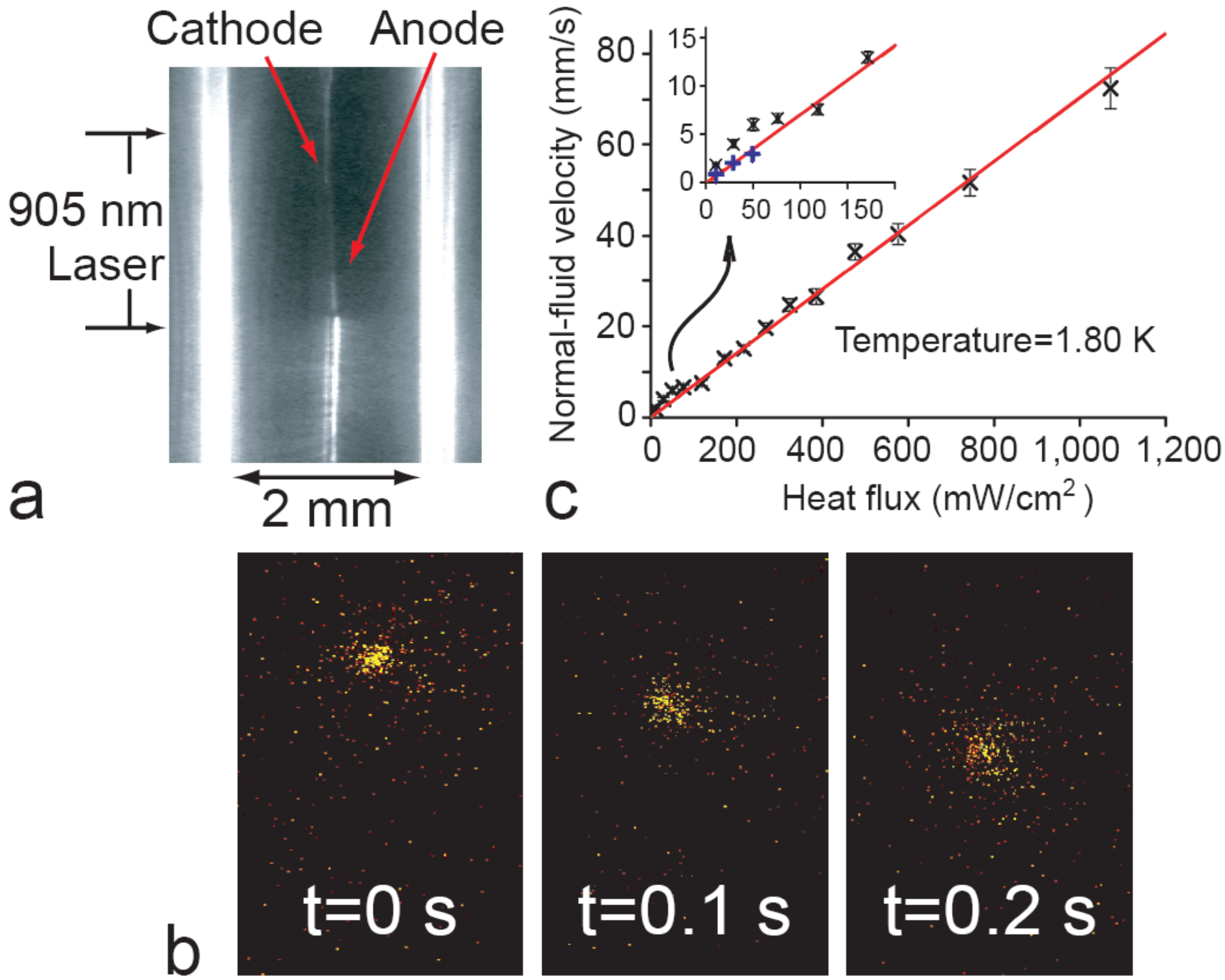

Figure 4: **Images and data obtained in the cluster tracking experiment**. **a,** An image of the square glass tube is shown. A pulsed laser at 905 nm illuminates the region of the needle apexes. **b,** Fluorescent images of a cluster of helium molecules taken at 0 s, 0.1 s and 0.2 s after the cluster was created are shown. The heat flux is 119 mW/cm$^2$. The temperature is 1.80 K. The exposure time for each image is 20 ms. **c,** The black crosses show the measured normal-fluid velocity as a function of heat flux at 1.80 K. Error bars represent the error in the linear fit to determine the normal-fluid velocity as discussed in the text. The red line is the theory curve based on equation (1) in the text. The blue crosses in the zoom-in inset for small heat flux regime show the corresponding measured normal-fluid velocity divided by a factor of 2.

**Supplementary Information**

**Section 1 - Description of the helium molecule imaging technique**

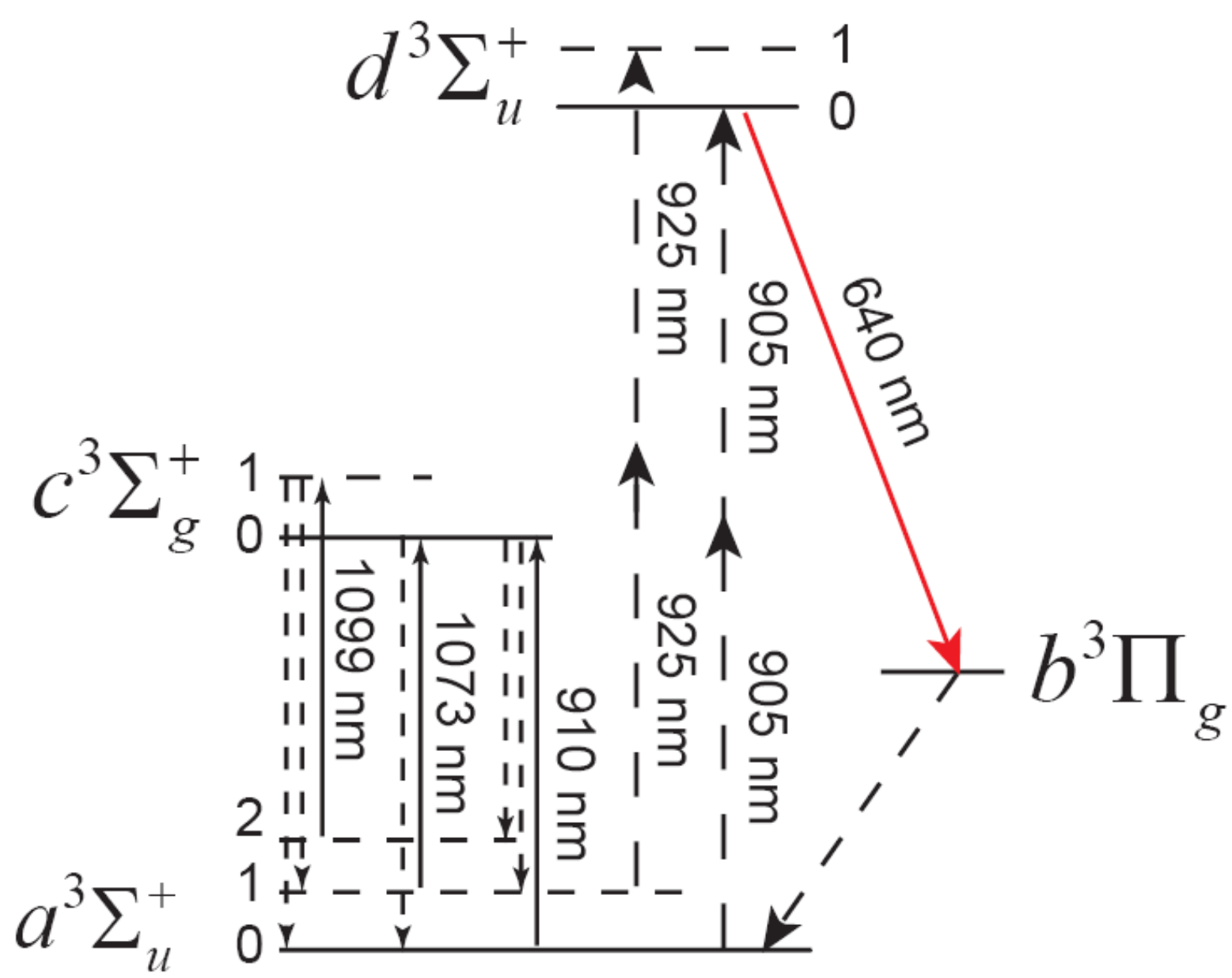

Figure S1: **Energy level diagram and optical transitions used to image the $He_2^*$ triplet molecules.** The levels labelled by 0, 1 and 2 for each electronic state are the vibrational levels of the corresponding state.

Two infra-red photons at 905 nm can excite a helium molecule from its triplet ground state $a^3\Sigma_u^+$ to the excited electronic state $d^3\Sigma_u^+$. Calculations of the branching ratios indicate that about 10% of the excited molecules in $d^3\Sigma_u^+$ state will decay to the $c^3\Sigma_g^+$ state, while the remaining 90% will decay to the $b^3\Pi_g$ state, emitting detectable red photons at 640 nm. Molecules in both the $c^3\Sigma_g^+$ and $b^3\Pi_g$ states then decay back to the $a^3\Sigma_u^+$ state, and the process can be repeated (see Fig. S1). However, the molecules falling to excited vibrational levels ($a$(1) and $a$(2) levels) of the $a^3\Sigma_u^+$ state are trapped in these long-lived off-resonant levels and are lost for subsequent cycles. To recover the molecules in the excited vibrational levels of the $a^3\Sigma_u^+$ state, continuous diode lasers at 1073 nm and 1099 nm can be used to drive the molecules from the *a*(1) state to the *c*(0) state and from the *a*(2) state to the *c*(1) state, respectively. Molecules in the *c*(0) and *c*(1) states predominantly decay to the *a*(0) state. In the molecule tagging experiment, on the other hand, the first vibrational level *a*(1) of the $a^3\Sigma_u^+$ state is used. A pump laser pulse at 910 nm can be used to drive the population from the triplet ground state *a*(0) to the excited state *c*(0), where the molecules will decay to the *a*(1) vibrational level roughly 4% of the time in a few nanoseconds (see Fig. S1). A probe laser pulse at 925 nm can then be used to image only the *a*(1) molecules at given delay time by driving the molecules from *a*(1) to the *d* state and inducing 640 nm fluorescence via *d* to *b* transition. In the experiment, molecules created by field emission initially occupy the *a*(0), *a*(1), and *a*(2) states. To prepare a pure population of *a*(0)-state molecules for tagging and eliminate background signal for selective imaging, the continuous diode lasers at 1073 nm and 1099 nm are used to illuminate the glass counterflow channel for 5 ms before each pump-probe trial.

**Section 2 - Description of the pulsed lasers**

The pulsed lasers we used in the molecule tagging experiment are Nd:Yag pumped Optical Parametric Oscillators (EKSPLA Model NT342/1). The lasers produce 4 ns pulses with a linewidth less than 5 $cm^{-1}$ and their output wavelengths can be tuned from 420-2300 nm. They have a maximum pulse energy of 9 mJ above 700 nm, and the repetition rate can be changed discretely from 0.1 Hz to 10 Hz. In the cluster tracking experiment, the pulsed laser at 905 nm is an OPO-based laser system using a diode-pumped Nd:Yag with a fixed repetition rate of 500 Hz. Its intensity in the experiment is about 300 $\mu$ J/cm$^2$ per pulse.

**Section 3 – Image acquisition**

We obtain images with an intensified CCD camera (Princeton Instruments PIMAX). A notch interference filter, centered at 640 nm, is placed in front of the camera to prevent any laser light or unwanted fluorescence from reaching the camera. The photocathode of the camera can be gated on and off externally. For images taken in the molecule tagging experiment, the gate signal of the camera is synchronized to each probe laser pulse, and the photocathode is gated on for only 6 $\mu s$ so as to minimize the dark current. The fluorescent light from a line of tagged helium molecules induced by a probe laser pulse is recorded on the CCD. To achieve high image quality, at each given pump-probe delay time, up to 40 images were superimposed. In the cluster tracking experiment, the gate signal (6 $\mu s$ duration) is synchronized to the pulsed laser at 905 nm whose repetition rate is 500 Hz. The exposure time (shutter open time) for each image is 20 ms, during which the molecules are exposed to 10 laser pulses.